\documentclass[paper,11pt]{JHEP}
\usepackage[centertags]{amsmath}
\usepackage{amsfonts} \usepackage{amssymb} \usepackage{amsthm}
\usepackage{graphicx}
\usepackage{cite}
\newcommand{\bra}{\langle}
\newcommand{\ket}{\rangle}

\def\one{{\rm 1\kern -.9mm l}}                             %

\def\beq{\begin{equation}}
\def\eeq{\end{equation}}
\def\beqa{\begin{eqnarray}}
\def\eeqa{\end{eqnarray}}
\newcommand{\eqa}{\begin{eqnarray}}
\newcommand{\ena}{\end{eqnarray}}

\newcommand{\Z}{\mathbb{Z}}

\newcommand{\C}{\mathbb{C}}

\newcommand{\cH}{\mathcal{H}}
\newcommand{\cJ}{\mathcal{J}}
\newcommand{\cN}{\mathcal{N}}
\newcommand{\oo}{\mathcal{O}}
\newcommand{\D}{\Delta}

\title{Anomalous dimensions of spinning operators\\ from conformal symmetry}
\author{ Ferdinando Gliozzi \\
Dipartimento di Fisica, Universit\`a di Torino\\
and Istituto Nazionale di Fisica Nucleare - sezione di Torino \\
Via P. Giuria 1, I-10125 Torino, Italy}

\abstract{
  We compute, to the first non-trivial order in the $\epsilon$-expansion of
  a perturbed  scalar field theory,
the anomalous dimensions of an infinite class of primary operators
with arbitrary spin $\ell=0,1,\dots$, including as a  particular case
the weakly broken higher-spin currents,
using only constraints from conformal symmetry.
Following the bootstrap philosophy, no reference is made to
any Lagrangian, equations of motion or coupling constants. Even the
space dimensions $d$ are left free. The interaction is implicitly turned
on through the local operators  by letting them acquire
anomalous dimensions. 

When matching certain four-point and  
five-point functions  with the  corresponding
quantities of the free field theory in the  $\epsilon\to 0$ limit, no free parameter remains. It turns out that only the expected discrete $d$ values are
permitted and the ensuing anomalous dimensions  reproduce known
results for the weakly broken higher-spin currents and provide new
results for the other spinning operators.}
\begin{document}

\section{Introduction}
\label{sec:intro}

The conformal bootstrap approach 
\cite{Ferrara:1973yt,Polyakov:1974gs,Belavin:1984vu},  in its modern
revival\cite{Rattazzi:2008pe,
Rychkov:2009ij, Rattazzi:2010gj,Poland:2010wg,ElShowk:2012ht,Liendo:2012hy,
Pappadopulo:2012jk,ElShowk:2012hu,Gliozzi:2013ysa, Gaiotto:2013nva,
 El-Showk:2014dwa,
Beem:2013qxa,Nakayama:2014yia,Nakayama:2014sba,Gliozzi:2014jsa,Chester:2014fya, 
Kos:2014bka,Chester:2014gqa,Beem:2014zpa,Simmons-Duffin:2015qma, 
Bobev:2015vsa,Kos:2015mba,Bobev:2015jxa,Gliozzi:2015qsa,
Beem:2015aoa,Nakayama:2016jhq,Kos:2016ysd,Nakayama:2016cim,
Gliozzi:2016cmg,Esterlis:2016psv,El-Showk:2016mxr,Hikami:2017hwv,Li:2017agi},
is the latest of  a set of powerful tools applied to the study of critical
systems in different dimensions. One of the most surprising features
of this approach is that some of the fundamental
properies characterizing the critical systems, like the scaling
dimensions of local operators and their operator product expansion
(OPE) coefficients are (numerically) obtained under the only
assumption that these systems are described by a conformal field
theory (CFT), with no reference to Lagrangians, coupling constants or
equations of motion.  On the contrary in all the  conventional
approaches, like  $\epsilon$-expansions \cite{Wilson:1973jj}  or Monte Carlo
calculations, the choice of a specific Lagrangian is mandatory. 
This intriguing difference between conventional and bootstrap approaches
calls for an analytical explanation. 

A first step in this direction was taken in
Ref.\cite{Rychkov:2015naa} in the context of
$\phi^4$ theory in $4-\epsilon$ dimensions. It was shown that the
anomalous dimensions of scalar operators at the first non-trivial
order in the $\epsilon$-expansion can be computed under the following
three axioms: (I) The Wilson-Fisher (WF) fixed point is described by a
CFT. (II) in the $\epsilon\to0$ limit the vacuum expectation values of
$n$-point functions approach those of the free field theory. (III) At
the WF fixed point $\phi^3$ is a descendant of $\phi$ as a consequence
of the equation of motion. Such an approach has been generalized in
various ways in \cite{Basu:2015gpa, Nii:2016lpa,Hasegawa:2016piv, 
  Bashmakov:2016uqk,Gliozzi:2016ysv,Liendo:2017wsn,Gliozzi:2017hni,
  Codello:2017qek}. In particular in  \cite{Gliozzi:2016ysv} it was shown that the axiom (III)
is redundant and, in the spirit of bootstrap, no equations of motion nor 
any kind of Lagrangians were assumed. How to turn on the interaction
without resorting to any interaction term or coupling constant? 
 The recipe is very simple: one looks for a  consistent smooth conformal 
deformation of  a free theory in $d-\epsilon$  
where the scaling dimensions $\D_\oo$ of the operators $\oo$ of
the deformed theory differ from those $\oo_f$ of the free theory with the same
quantum numbers by  corrections which can be expanded in
powers of $\epsilon$
\beq
\D_\oo=\D_{\oo_f}+\gamma_\oo=\D_{\oo_f}+\gamma^{(1)}_\oo\epsilon+\gamma^{(2)}_\oo\epsilon^2+\dots
\label{eq:anomalous}
\eeq
where the $\gamma_\oo$'s are the anomalous dimensions. A stringent requirement
is that there is a one-to-one mapping between the local
operators of the free and the interacting theory. It turns out
that this deformation can be consistently made only for some special
values of the space-time dimensions $d$.  These coincide with the
upper critical dimension $d_m=2m/(m-2)$
(integer or rational \cite{Gracey:2017okb})  where, in the standard RG approaches to a
scalar  theory perturbed with a $\phi^m$ potential, the WF fixed point
is defined as a non-trivial zero of the $\beta$-function in
$d=d_m-\epsilon$. 

Matching suitable four-point functions in the
$\epsilon\to0$ limit with the free theory data  yields
at once the anomalous dimensions of $\phi^p$ for any integer $p$ at
the first non-trivial order, while so far no information was obtained
on the spectrum of spinning operators within such a method. There are
other approaches that provide it for the special case of
weakly broken higher-spin (HS) currents. For instance, from a Mellin  space
representation of the
four-point, the anomalous dimensions of these currents and  other scalar operators up to
$\epsilon^3$   \cite{Gopakumar:2016wkt,Gopakumar:2016cpb,Dey:2016mcs}
and even higher order \cite{Dey:2017oim} have been obtained. Similarly,
in a study relying in a standard Lagrangian description of conformal
theories with weakly broken HS symmetry, the leading order
anomalous dimensions of these currents have been determined from
the classical non-conservation equations \cite{Skvortsov:2015pea,Giombi:2016hkj}. WF
fixed points with $O(N)$ symmetry have been studied in the large
global charges limit in \cite{Hellerman:2015nra}. In another study 
relying instead in the conformal bootstrap \cite{Alday:2016njk,Alday:2016jfr}, 
the spectrum of these broken currents,
at the leading order, follows from crossing symmetry and the exact
conservation of the stress tensor.
 
The aim of this paper is to compute, at the leading order in
$\epsilon$, the anomalous dimensions of a much
larger class of primary operators $\oo_{p,\ell}$ of arbitrary spin
$\ell$, namely a set of
those that in the Gaussian limit can be written in terms of symmetric
traceless combinations of $\ell$
derivatives and $p+1$ copies of the scalar field $\phi$. Their scaling 
dimensions can be written as
\beq
\D_{p,\ell}=(p+1)\frac{d-2}2+\ell+\gamma_{p,\ell}\,,
\label{eq:spindim}
\eeq
with 
$\gamma_{p,\ell}=\gamma_{p,\ell}^{(1)}\epsilon+\gamma_{p,\ell}^{(2)}\epsilon^2+\dots$.
The special case $p=1$ describes the HS conserved currents
which are no longer conserved  (for $\ell\not=2$)  at the WF fixed
point, as they acquire an anomalous dimension; they are called sometimes weakly broken HS currents.  On the contrary the stress tensor turns out to be 
exactly conserved in the present approach, as expected.    

All our calculations rely  in a unique group-theoretical mechanism,
the conformal multiplet recombination first pointed out in
\cite{Rychkov:2015naa}. Here 
we apply it in the form developed in \cite{Gliozzi:2016ysv,Gliozzi:2017hni}. 
Schematically, taking $d\sim4$
for simplicity, the argument runs as follows.
 Consider the following fusion rule of the Gaussian theory
\beq
    [\oo_{p,\ell}]\times[\phi^p_f]={\sf c}_1 [\phi_f]+{\sf c}_2 [\phi^3_f]+\dots,
 \label{eq:freefusion}   
\eeq 
where  $[\oo]$ denotes a (suitably normalized) primary, the 
${\sf c}_i$'s are OPE coefficients and the dots contain all  other
primaries which do not participate in the game. Using the equation above, the
four-point function $\bra\oo_{p,\ell}\,\phi_f^p\,\oo_{p,\ell}\,\phi_f^p\ket$
can be written as
\beq
 \bra\phi_f^p\,\oo_{p,\ell}\,\oo_{p,\ell}\,\phi_f^p\ket={\sf c}_1^2
\sum_{\alpha\in\cH_{\phi_f}}
\frac{\bra\phi_f^p\oo_{p,\ell}\vert\alpha\ket\bra\alpha\vert\oo_{p,\ell}\phi_f^p\ket}{\bra\alpha\vert\alpha\ket}+{\sf c}_2^2\sum_{\alpha\in\cH_{\phi_f^3}}
\frac{\bra\phi_f^p\oo_{p,\ell}\vert\alpha\ket\bra\alpha\vert\oo_{p,\ell}\phi_f^p\ket}{\bra\alpha\vert\alpha\ket}
+\dots
\label{fourf}
\eeq
$\alpha\in\cH_{\oo}$ denotes a state of the conformal multiplet
generated by the primary $\oo\vert0\ket=\vert\oo\ket $ and
normalized 
so that $\bra\oo\vert\oo\ket=1$.
At the WF fixed point,
because of the mentioned recombination, $[\phi^3]$ is
absorbed by $[\phi]$ as a sub-representation and the  fusion rule becomes 
 \beq
[\oo_{p,\ell}]\times[\phi^p]=({\sf c}_1+O(\epsilon)) [\phi]+\dots.
\eeq 
Thus in the corresponding four-point function the second 
sum disappears while in the first sum  new terms show up because
now the conformal multiplet $[\phi]$ becomes long:
\beqa
\bra\phi^p\,\oo_{p,\ell}\,\oo_{p,\ell}\,\phi^p\ket=&
\left({\sf c}_1^2+O(\epsilon)\right)
\Big(\sum_{\alpha\in\cH_{\phi_f}}
\frac{\bra\phi^p_f\,\oo_{p,\ell}\vert\alpha\ket\bra\alpha\vert\oo_{p,\ell}\,
\phi^p_f\ket}{\bra\alpha\vert\alpha\ket}+\nonumber\\&
\sum_{\beta\in\cH_\chi}\frac{\bra\phi^p\oo_{p,\ell}\vert\beta
\ket\bra\beta
\vert\oo_{p,\ell}\,\phi^p\ket}{\bra\beta\vert\beta\ket}\Big)+\dots
\label{fouri}
\eeqa
$\chi$ is a  descendant of $\phi$ with scaling dimension
$\D_\chi=\frac d2+1+O(\epsilon^2)$.
The  norm of $\vert\chi\ket$  depends on that of $\vert\phi\ket$ and goes to zero in the $\epsilon\to0$ limit (for more details see next section). Actually the
${\sf  c}^2_2$ of the free theory is replaced by
\beq
    {\sf  c}^2_2\to \,{\sf  c}^2_1\frac{\bra\phi^p\,\oo_{p,\ell}\vert\chi\ket
      \bra\chi\vert\oo_{p,\ell}\,\phi^p\ket}{\bra\chi\vert\chi\ket}
\label{eq:c2toc1}
\eeq

In the $\epsilon\to0$ limit  $\chi$  is
indistinguishable from $\phi_f^3$,
hence the two expansions (\ref{fourf}) and (\ref{fouri}) must coincide.
We find an infinite set of matching conditions, one for each pair
of integers $p=1,2,\dots$ and $\ell=0,1,\dots$,

\beq
\lim_{\epsilon\to0}\frac{\left[(\gamma_\phi+\gamma_{\phi^p}-\gamma_{p,\ell})
(\gamma_\phi+\gamma_{\phi^p}-\gamma_{p,\ell}+2-d-2\ell)\right]^2}
 {4d\,(d-2)\,\gamma_\phi}=\left(\frac{{\sf c}_2}{{\sf c}_1}\right)^2\,.
\label{eq:match}
\eeq 
In this form such constraints are valid not only for $d$ near 4, but also for
$d$ near 3, the upper critical dimension of a $\phi^6$ perturbation and, more generally, for the fractional $d=2k/(k-1)$ corresponding to the upper critical dimension of the multicritical points of higher order, associated with
$\phi^{2k}$ potentials.

For $\ell=0$ the equations  above  coincide with those found in
\cite{Gliozzi:2016ysv,Gliozzi:2017hni} and used to compute 
$\gamma_{\phi^p}$ to the first order 
for any integer $p>1$ and to the second order for $p=1$. Here we extend 
the calculation to any spin
$\ell$.  In the particular case of weakly broken higher-spin currents, i.e. $p=1$, we have ${\sf c}_2=0$ and
these matching conditions tell us  simply that
$\gamma_{1,\ell}^{(1)}=0$, a well known property of the WF fixed point
at $d=4-\epsilon$ and at $d=3-\epsilon$. In order to calculate the first
non-vanishing term of  
$\gamma_\ell\equiv\gamma_{1,\ell}$ we found it useful to study certain five-point 
functions which lead to the right answer in a surprisingly simple way.
We found indeed the following new set of matching conditions
\beq
\lim_{\epsilon\to0}\frac{(\gamma_\ell-2\gamma_\phi)(d-4+2\ell+\gamma_\ell-
2\gamma_\phi)(d-2+2\ell)}
{2d(d-2)\gamma_\phi}=m-1\,.
\label{eq:match5}
\eeq
Here $m$ is an arbitrary integer $m>2$ and $d=d_m-\epsilon$, where
\beq
d_m=\frac{2m}{m-2}
\label{eq:upperd}
\eeq
 is the upper critical dimension for a perturbing potential $\phi^m$.
In particular these matching conditions give at once, at the first 
non-vanishing order, the anomalous dimensions of the weakly broken HS
currents for any $\ell$ in $d-\epsilon$ space-time dimensions with $d=3,4,6$. 
According to our definitions, $\gamma_{\ell=0}\equiv\gamma_{\phi^2}$. In such
a case
these equations allow to compute the $\epsilon^2$  contribution of 
$\gamma_{\phi^2}$ in the multicritical cases. This was not possible with the first set (\ref{eq:match}) of constraints. 

Admittedly, owing to the huge amount of tensor structures involved, 
no much is known about the properties of the conformal block expansions 
of the four-point function (\ref{fouri}) for general $\ell$. Much less
is known about  similar expansions of the five-point. 
However  we are only interested in the couplings of the
external operators to the scalar $\chi$. Here only a single tensor
structure appears and all calculations simplify dramatically.

This paper is organized as follows. In section 2 we describe a simple 
algebraic method to classify all the possible representations of the
conformal group in $d$ dimensions admitting a primary descendant in 
the case of symmetric traceless representations of the rotation 
subgroup $SO(d)$. In section 3 we collect some useful properties of
primary operators with spin, in section 4 we write  the general
matching conditions and their consequences. The body of the paper
includes also some conclusions. Appendix A contains some 
formulas of the free field theory which are useful to compute the 
OPE coefficients ${\sf c}_1$ and ${\sf c}_2 $.
\section{Null States}
\label{sec:null}
The Euclidean conformal group $SO(d+1,1)$ admits representations that are reducible but not completely reducible, i.e. they can have an invariant subspace whose complement is not invariant. The states of this invariant subspace have  null norm
(null states) and the corresponding multiplet is said to be long.
The bosonic free field theory does not have this kind of representations (the corresponding conformal multiplets are short). The Wilson-Fisher fixed points can be seen as the points where some multiplets recombine to form a long multiplet \footnote{Actually this multiplet does not contain exactly null states, but states whose norm goes to zero as their scaling dimension approaches the free field value.}.

It is instructive to investigate the mechanism of appearance of null states
operating with the  Lie algebra of the conformal group. A  standard basis is
$D,P_\mu,K_\mu,J_{\mu\,\nu}$, $(\mu,\nu=1,\dots d)$,  which generate respectively
the dilatation,  the translations, the special conformal
transformations and the $SO(d)$ rotations. Applying this algebra to a state
$\vert\D,\ell\ket=\lim_{\epsilon\to0}\oo_{\D,\ell}(x)\vert0\ket$, where $\oo_{\D,\ell}$ is an arbitrary local operator and $\vert0\ket$ a conformally invariant vacuum, generates a
whole representation of the conformal group. A state with minimal
$\D_{min}=\D'$ is said primary, all  other states, where $\D=\D'+n$, $(n=1,2\dots)$ are descendants.
A primary state is annihilated by the $K_\mu$'s. If $\vert\D',\ell'\ket$ is the only state which is annihilated by the  $K_\mu$'s the representation is irreducible, while if there is a primary descendant, i.e. a descendant state with
$K_\mu\vert\D'+n,\ell\ket=0$, this state has null norm and generates an invariant subspace  of null states.

The conformal multiplets generated by the fusion of two scalars correspond to symmetric, traceless tensors.
The complete list of the primary descendants that could appear in the OPE of
two scalars can be obtained in various ways. One point of attack follows form
the study of the conformal blocks $G_{\D,\ell}(u=\frac{x_{12}^2x_{34}^2}{x_{13}^2x_{24}^2},v=\frac{x_{23}^2x_{14}^2}{x_{13}^2x_{24}^2})$ contributing to a four-point function of scalars $\bra\oo_1(x_1)\oo_2(x_2)\oo_3(x_3)\oo_4(x_4)\ket$. These are eigenfunctions of the Casimir invariants $C_2,C_4,\dots$ that can be formed with the Lie generators 
\beq
C_i G_{\D,\ell}(u,v)=c_i(\D,\ell)\,G_{\D,\ell}(u,v)\,,
\label{eq:conformalblock}
\eeq
where the first two Casimir eigenvalues are, respectively, 
\begin{align} \label{eq:casimirequations}
  c_2(\Delta, \ell) &= \frac{1}{2}\Delta ( \Delta -d ) +
  \frac{1}{2}\ell(\ell + d -2)\, ,\\
  c_4(\Delta, \ell) &= \Delta^2 ( \Delta -d )^2 +
  {1 \over 2} d (d-1) \Delta (\Delta -d)+ 
\ell^2 (\ell+d -2)^2 + {1 \over 2} (d-1)(d-4)\ell (\ell+d-2) \, . \nonumber
\end{align}
Higher order Casimir invariants are redundant when considering  symmetric, traceless tensors.
These conformal blocks can be written as sum of poles in $\D$. As Eq. (\ref{fouri}) suggests, poles occur at special $\D$'s where a descendant state becomes 
null. 
The location of poles, their residues as well as the remaining entire function can be computed \cite{Kos:2014bka} solving  the Casimir differential equations.
Ref. \cite{Penedones:2015aga} uses instead a group-theoretical method, looking for all primary descendants that arise when the dimension $\D'$ of the primary
varies. In both cases one finds three sequences of poles, as shown in table 1.

Here we give a two-line algebraic derivation of such a table. Let
$\vert\D',\ell'\ket$ be a primary state and $\vert\D=\D'+n,\ell\ket$ a primary descendant. Since they
belong to the same representation, they must share the eigenvalues of the Casimir invariants, namely
\beq
c_2(\D',\ell')=c_2(\D,\ell)~,\,c_4(\D',\ell')=c_4(\D,\ell)\,.
\eeq
As first pointed out in \cite{Gliozzi:2017hni}, the solutions of these two
algebraic equations yield the complete list of   primaries
admitting a primary descendant, reproduced in table 1. End of the proof\footnote
{As already mentioned, such a list applies to primaries that belong to the
traceless symmetric representations of $SO(d)$. For more general
representations higher order Casimir invariants are necessary.}.
\begin{table}
\centering
\begin{tabular}{c|c|c|cl}
Parent primary &\multicolumn{2}{c}{Primary descendant}\\
\hline
$\D'_k$  & $\Delta_k$ & $\ell$  \\
\cline{1-4}
$1-\ell'-k$ & $1-\ell +k$ & $\ell'+k$    & \quad $k=1,2,\dots$\\
${d \over 2}-k$  & ${d \over 2}+k$  & $\ell'$     & \quad $k=1,2,\dots$\\ 
$d+\ell'-k-1$ & $d+\ell+k-1$ & $\ell'-k$   & \quad $k=1,2,\dots, \ell $
\end{tabular}
\caption{List of solutions of the algebraic system
  (\ref{eq:casimirequations}). The $\D'_k$'s in the first column give the positions of  poles of a generic conformal block defined in (\ref{eq:conformalblock}). The second  and third columns list the scaling dimension $\D_k$ and the spin
  $\ell$ of the primary descendant belonging to the representation generated by the parent primary $\vert\D',\ell'\ket$.}
\end{table}

If $d$ is an even integer, some of the above solutions coalesce, giving rise
to double poles in the conformal blocks \cite{Penedones:2015aga}. In this paper we study theories in $d-\epsilon$ dimensions, so we do not need to consider these exceptional cases.

Among the primaries admitting a primary descendant, the one with $\D'=\frac d2-1$ and $\ell'=0$ has the same quantum numbers of a scalar free field $\phi_f$ in $d$ space-time dimensions. Is there any restriction in a unitary theory
assuming this  \underline{\sl to be} a free field? The answer is no,
of course, but since this fact is not always clearly stated in
literature, we quote here a theorem due to Weinberg \cite{Weinberg:2012cd}:
In a scale-invariant relativistic theory any local scalar operator $\psi(x)$
having scaling dimension $\D_\psi=\frac d2-1$ is necessarily a free field.
 We have again a  proof of few lines \footnote{The Weinberg proof is for $d=4$ and also refers  to a class of spinning operators.
  Here we give its obvious generalization to any $d$ for scalar operators.}. Here, having chosen  Euclidean signature, we have to replace Lorentz
 invariance with $SO(d)$ invariance. The latter implies
 \beq
 J_{\mu\,\nu}\,G(z)=0\,,
 \label{eq:sodinvariance}
 \eeq
 with
 \beq
 G(x-y)\equiv\bra0\vert\psi^\dagger(x)\,\psi(y)\vert0\ket~,~~~J_{\mu\,\nu}\equiv-i z^\mu
 \frac\partial{\partial z_\nu}+i z^\nu
 \frac\partial{\partial z_\mu}\,.
 \eeq
 Using such a differential representation one can derive the identity
 \beq
 J^2\equiv J_{\mu\,\nu}J^{\mu\,\nu}=-2z^2\partial^2-2D(D-i(d-2))\,,
 \label{eq:j2}
 \eeq
 where $D=-iz^\rho\frac\partial{\partial z_\rho}$ is the generator of the
 dilatations. If $\psi$ has scaling dimensions $\frac d2-1$, in a scale-invariant theory 
 \beq
 D\,G(z)=i(d-2)G(z)\,,
 \label{eq:scaling}
 \eeq
 so Eq.s (\ref{eq:sodinvariance}), (\ref{eq:j2}) and (\ref{eq:scaling}) together  give $\partial^2\,G(z)=0$.
 Operating again with $\partial^2$, it follows trivially that, assuming
 unitarity,
 \beq
 \bra0\vert\left(\partial_x^2\psi(x)\right)^\dagger\,
 \partial^2_y\psi(y)\vert0\ket=0\,~\Rightarrow ~
 \partial^2_y\psi(y)\vert0\ket=0\,.
 \eeq
 But any local operator that annihilates the vacuum must vanish, so
 \beq
 \partial^2_x\psi(x)=0\,.
 \label{eq:freefield}
 \eeq
 It is interesting to note that  in this way the equation of motion of  a
 free field  has been obtained in a purely group-theoretical manner,
 with no reference to Lagrangians or to any sort of dynamical principles. 
 
 Ref. \cite{Penedones:2015aga} obtained the explicit form of the polynomials
 in the  Lie algebra elements which generate null states when applied to the primaries listed in the first column of
 table 1. That transforming the primary state
 $\vert\frac d2-1,0\ket$ into the corresponding null state of the second
 column of the table is
 $P^2 \vert\frac d2-1,0\ket\equiv-\partial^2
 \vert\frac d2-1,0\ket=\vert\frac d2+1,0\ket$
 which in view of the above theorem is actually zero.
 Therefore the conformal multiplet generated by a bosonic free field $\phi_f$ is short. Turning on the interaction by letting $\phi$ acquire anomalous
 dimensions implies that this descendant is no longer zero. However, strictly speaking, it is no longer a null state; in fact its norm is \cite{Penedones:2015aga}
 \beq
 \bra 0,\D\vert K^2\,P^2 \vert\D,0\ket=8d \D(\D-\D_{\phi_f})\,,
 \label{eq:chi}
 \eeq
 as a consequence of  the commutation relations
 \beq
 [K_\mu,P_\nu]=2i\left(\eta_{\mu\nu}D-J_{\mu\nu}\right)\,,
 \eeq
 combined with the constraints
 \beq
 K_\mu\vert\D,0\ket= J_{\mu\nu}\vert\D,0\ket=0\,,~~\bra0,\D\vert P_\mu=\bra0,\D\vert J_{\mu\nu}=0\,,~~~\bra0,\D\vert\D,0\ket=1\,.
 \eeq
 Putting $\D=\D_\phi=\D_{\phi_f}+\gamma_\phi$, we see that this state becomes null
 when it approaches the free field theory, where it disappears.
 This fact has an important role in the study of the possible smooth deformations of a free theory. As mentioned in the introduction, we say that a free field theory in $d$ dimensions admits
 a smooth deformation in $d-\epsilon$ if there is a one-to-one mapping to
 another CFT in which any local operator $\oo_f$ of the free theory corresponds to an operator $\oo$ of the deformed theory with the same spin and
 $\lim_{\epsilon\to0}\D_\oo=\D_{\oo_f}$. For a general $d$ such a deformation does not exists because $\partial^2\phi$ does not have a corresponding operator in the free theory. There are however  special values of $d$,  where there is a scalar $\phi^{m-1}_f$ having the same scaling dimensions of the vanishing operator
 $\partial^2\phi_f$ which restores the one-to-one correspondence. Thus from
 the degeneracy condition $(m-1)\D_{\phi_f}=\D_{\phi_f}+2$  the 
upper critical dimension (\ref{eq:upperd}) follows.

 \section{Spinning Operators}
 \label{sec:spin}
 In this section we use the methods and the notations of  
 \cite{Penedones:2015aga} to derive some useful properties of local operators
 belonging to  symmetric and traceless representations of $SO(d)$.
 A primary operator of spin $\ell$ is denoted by
 \beq
 \oo(x,z)=\oo^{\mu_1\dots\mu_\ell}(x)z_{\mu_1}\dots z_{\mu_\ell}\,,
 \label{eq:operatorwithspin}
 \eeq
 where $z_\mu$ is a vector of $\C^d$ that can be consistently assumed to
 satisfy $z\cdot z=0$ as a consequence of the null traces of the tensor
 (\ref{eq:operatorwithspin}). One can recover the tensor indices by applying recursively the Todorov differential operator defined as\cite{Dobrev:1976vr}
 \beq
 D^\mu_z\equiv\left(\nu+z\cdot\frac\partial{\partial z}\right)
 \frac\partial{\partial z_\mu}-\frac12 z^\mu\frac{\partial^2}{\partial z\cdot\partial z}\,.
 \label{eq:todorov}
 \eeq
 where $\nu=\frac d2-1$.

 The two-point function of a canonically normalized primary of spin $\ell$ and scaling dimensions $\D$ is
 \beq
 \bra \oo(x_1,z_1)\oo(x_2,z_2)\ket
 =\frac{\left(z_1^\mu I_{\mu\nu}(x_{12}) z_2^\nu\right)^\ell}{(x_{12}^2)^\D}
 \eeq
 with $x_{ij}\equiv x_i-x_j$ and $I_{\mu\nu}(x)\equiv
 \delta_{\mu\nu}-2x_\mu  x_\nu/x^2$.

 The three-point of the spinning primary $\oo_1(x_1,z)$ with
 two scalar primaries $\oo_2(x_2)$ and $\oo_3(x_3)$ reads
 \beq
 \bra \oo_1(x_1,z)\oo_2(x_2)\oo_3(x_3)\ket=
 \frac{{\sf c}_{123}\,\,\left(z\cdot y\right)^\ell}{\left(x_{12}^2\right)^{\frac{\D_1+\D_2-\D_3-\ell}2}
   \left(x_{23}^2\right)^{\frac{\D_2+\D_3-\D_1+\ell}2}
   \left(x_{31}^2\right)^{\frac{\D_3+\D_1-\D_2-\ell}2}}\,,
 \label{eq:threepoint}
 \eeq
 with $ y^\mu\equiv \frac{x_{31}^\mu}{x_{31}^2}-\frac{x_{12}^\mu}{x_{12}^2}$. 
 We can infer the OPE
 \beq
 \oo_1(x,z)\,\oo_2(0)={\sf c}_{123}
 \frac{(-x\cdot z)^\ell}{\left(x^2\right)^{\frac{\D_1+\D_2-\D_3+\ell}2}}\oo_3(0)+\dots\,,
 \label{eq:OPE}
 \eeq
 indeed inserting this OPE in the LHS of (\ref{eq:threepoint}) with
 $x^\mu=x_1^\mu$, $x_2^\mu=0$ and  $x^2\ll x^2_3$  yields the RHS.

 We want to exploit this OPE to obtain two important and useful properties of
 the conserved spinning operators (see also \cite{Costa:2011mg}). Within these notations the conservation
 law of $\cJ_\ell(x,z)\equiv\oo_1(x,z)$ reads
 \beq
 \frac\partial{\partial x}\cdot D_z\,\cJ_\ell(x,z)=0\,.
 \label{eq:conservation}
 \eeq
 Applying this differential operator to both sides of  (\ref{eq:OPE}) gives
 \beq
 {\sf c}_{\cJ\oo_2\oo_3}\left(d-4+2\ell\right)\left(\D_\ell+\D_2-\D_3-d+2-\ell\right)=0\,.
 \eeq
 For $\D_2=\D_3$ we obtain  
$\D_\ell=d-2+\ell$, the well known scaling dimension 
 of a conserved HS current. For $\D_2\not=\D_3$ the above equation 
tells us that the OPE coefficient must vanish:
 \beq
 \frac\partial{\partial x}\cdot D_z\,\cJ_\ell=0~ \& ~\D_2\not=\D_3\,
 \Rightarrow~{\sf c}_{\cJ\oo_2\oo_3}=0\,.
\label{eq:theorem}
 \eeq
 Such a theorem will be useful in our derivation of the anomalous dimensions of
 weakly broken HS currents.

 We write now the previous OPE in the form
 \beq
 \phi^p(0)\,\phi(x)={\sf c}_{p\,\ell}\frac{(x\cdot z)^\ell}
{\left(x^2\right)^{\frac{\D_{\phi^p}+\D_\phi-\D_{p,\ell}+\ell}2}} \oo_{p,\ell}(0,z)+\dots\,,
 \label{eq:OPE1}
 \eeq
 where $\oo_{p,\ell}$ is a spinning primary   with scaling dimensions
 $\D_{p,\ell}=(p+1)\D_{\phi_f}+\gamma_{p,\ell}$ in $d-\epsilon$ as in 
(\ref{eq:spindim}), $\phi$ is an  interacting scalar field and 
$\phi^p$ is normalized in such a way that
 \beq
 \bra\phi^p(x)\,\phi^p(0)\ket=\frac1{(x^2)^{\D_{\phi^p}}}\,.
 \eeq

A simple example of a spinning primary of the free field theory is
\beq
\oo_{2,2}^f=\cN\left[\phi_f\left(z\cdot\partial\phi_f\right)^2-
\frac{d-2}d\phi_f^2\left(
z\cdot\partial\right)^2\phi_f\right]\,,
\label{eq:free2,2}
\eeq
where $\cN$ is a normalization factor. Notice that there is an overall sign 
ambiguity in the definition of $\oo_{p,\ell}$ which entails a sign ambiguity
in the 
OPE coefficients.  The anomalous dimension of $\oo_{p,\ell}$ depends on ratios of these ${\sf c}$'s, so this sign ambiguity as well as the dependence on the normalization factor $\cN$ disappear.

The computed anomalous dimensions of this kind of spinning operators are explicitly written in Eq. (\ref{eq:spin2}).  
It is worth to stress that such a result is only valid for a {\sl primary} operator. For instance $[\phi^3]$ has a descendant of spin 2 which is degenerate 
with the above primary in the Gaussian limit, while its anomalous dimension 
is $\gamma_{\phi^3}$  (see Eq.(\ref{eq:solution0})), which is very different from (\ref{eq:spin2})\footnote{Notice that the spectrum of local operators of the free-field 
theory is highly degenerate at large $\ell$. This degeneracy is completely removed at the WF fixed point.}.

Using the explicit expression (\ref{eq:free2,2}) it is easy to evaluate ${\sf c}_{2,2}$, however this direct method would imply lengthy calculations for general ${\sf c}_{p,\ell}$. A much simpler approach is based on the following observation: all the $\oo_{p,\ell}$ are generated by the fusion rule
\beq
[\phi^p_f]\times[\phi_f]=\sqrt{p}[\phi_f^{p-1}]+\sum_{\ell=0}^\infty{\sf c}_{p,\ell}
[\oo_{p,\ell}^f]\,.
\label{eq:fusionspin}
\eeq
Appendix A describes a way to obtain these coefficients in the free field theory. The result is
\beq
{\sf c}^2_{p,\ell}=\frac{p(\nu)_\ell(2\nu)_\ell+(-1)^\ell(2\nu)_\ell(p\,\nu)_\ell}
{\ell!\left((p+1)\nu+\ell-1\right)_\ell}\,,
\label{eq:cpl}
\eeq
  where $(x)_y\equiv\frac{\Gamma(x+y)}{\Gamma(x)}$ is the Pochhammer symbol and $\nu=\frac d2-1$. In the interacting theory we have
\beq
\bra\phi^p\vert\oo_{p,\ell}\vert\phi\ket={\sf c}_{p,\ell}+O(\epsilon)\,.
\label{eq:cinteracting}
\eeq
In accordance with (\ref{eq:c2toc1}) we need to compute 
$\bra\phi^p\vert\oo_{p,\ell}\vert\chi\ket\equiv\bra\phi^p
\vert\oo_{p,\ell} P^2\,\vert \phi\ket$, where $\bra\phi^p\vert=
\lim_{x^2\to\infty}\bra0\vert\phi^p(x)(x^2)^{\D_{\phi^p}}$. Here
$\vert \chi\ket=\lim_{x\to0}P^2\phi(x)\vert0\ket$ and $\oo_{p,\ell}(x,z)$ is evaluated at $x^2=1$ and $z\cdot x=1$,  so that the associated three-point function coincides numerically with the OPE coefficient. Operating with 
$P^2=-\partial^2$ on both sides of (\ref{eq:OPE1}) it follows at once
 \beq
\bra\phi^p\vert\oo_{p,\ell}\vert\chi\ket={\sf M}_{p,\ell}\,
\bra\phi^p\vert\oo_{p,\ell}\vert\phi\ket
\eeq
with\footnote{Notice that {\sf M} is very similar, but not identical to the analogous quantity calculated in \cite{Penedones:2015aga}. There it referred to a 
primary descendant of a spinning operator that appears in the OPE of 
two scalars. Here the spinning primary is an external operator and
${\sf M}$ refers to $\chi$, the primary descendant of $\phi$.}
\beqa
{\sf M}_{p,\ell}=&\left(\D_\phi+\D_{\phi^p}-\D_{p,\ell}+\ell\right)
\left(\D_{p,\ell}-\D_\phi-\D_{\phi^p}-2+d+\ell\right)\nonumber\\
&=\left(\gamma_\phi+\gamma_{\phi^p}-\gamma_{p,\ell}\right)
\left(d-2+2\ell\right)+O(\epsilon^2)\,.
\label{eq:M}
\eeqa
In our study of certain five-point functions we also need  to know  
$\bra\chi\vert\cJ_\ell\vert\chi\ket\equiv
\bra\phi\vert\, K^2\cJ_\ell P^2\,\vert \phi\ket$. Repeating the above
calculation we find
\beq
\bra\chi\vert\cJ_\ell\vert\chi\ket={\sf N}_\ell\,
\bra\phi\vert\,\cJ_\ell\,\vert \phi\ket\,,
\eeq 
with
\beqa
{\sf N}_\ell=&{\sf M}_{1,\ell}\left(2\D_\phi-\D_{1,\ell}+2+\ell\right)
\left(\D_{1,\ell}-2\D_\phi+d-4+\ell\right)\nonumber\\
&=2\left(2\gamma_\phi-\gamma_{\ell}\right)
\left(d-2+2\ell\right)\left(\gamma_{\ell}-2\gamma_\phi+ d-4+2\ell\right)+O(\epsilon^3)\,.
\label{eq:N}
\eeqa
The denominator of (\ref{eq:c2toc1}) has been already computed in (\ref{eq:chi}); it reads
\beq
\frac1{\bra\chi\vert\chi\ket}\equiv\frac1{\bra\phi\vert\,K^2\,P^2\,
  \vert\phi\ket}=\frac{{\sf Q}}{\D_\phi-\D_{\phi_f}}=
\frac{{\sf Q}}{\gamma_\phi}\,,
\label{eq:denominator}
\eeq
with
\beq
    {\sf Q}=\frac1{4d(d-2)}+O(\epsilon)\,.
\label{eq:Q}    
\eeq
Equations (\ref{eq:M}), (\ref{eq:N}) and (\ref{eq:Q}) are the main ingredients to formulate the matching conditions discussed in the next section. 
\section{Matching Conditions}
\label{sec:match}

    The free field theory of a single real scalar is invariant under the $\Z_2$
    transformation $\phi_f\to-\phi_f$, therefore the fusion rule (\ref{eq:freefusion}) can be  only generalized  to those upper dimensions $d_{m}$ where the
    primary descendant $\phi_f^{m-1}$  is an odd power $2q+1=m-1$ of $\phi$
\beq
    [\oo_{p,\ell}^f]\times[\phi_f^p]={\sf c}_{\oo_{p,\ell}\phi^p\phi}[\phi_f]+
    {\sf c}_{\oo_{p,\ell}\phi^p\phi^{2q+1}}[\phi^{2q+1}_f]+\dots\,.
    \label{eq:freefusiong}
    \eeq
    The first OPE coefficients ${\sf c}_{\oo_{p,\ell}\phi^p\phi}\equiv{\sf c}_{p,l}$
    can be extracted from the fusion  $[\phi^p_f]\times [\phi_f]$ as we saw
    in the previous section (see Eq.(\ref{eq:cpl})). Similarly the second OPE
    coefficients ${\sf c}_{\oo_{p,\ell}\phi^p\phi^{2q+1}}\equiv{\sf c}_{q,p,l}$
    can be computed from the fusion   $[\phi^p_f]\times [\phi_f^{2q+1}]$.
    More specifically, as Eq.(\ref{eq:gexpansion}) shows, they are the 
coefficients of the following conformally invariant function
    \beqa
    \frac{p!(2q+1)!}{(q!)^2}u^{\nu(p+1)/2}\sum_{k=0}^{q+1}
\frac{v^{-k\nu}}{(k!)^2(p-q-k)!(q-k+1)!}=\nonumber\\
    \sum_{n=0}^{\infty}\sum_{\ell=0}^{\infty}{\sf c}_{n,q,p,\ell}^2
    G_{\nu(p+1)+2n+\ell,\ell}(u,v)\,,
    \label{eq:cq}
    \eeqa
    where  the $G_{\D,\ell}$'s are the conformal blocks
    contributing to the $\phi^p_f \phi_f^{2q+1}$ channel of the four-point
    function
    $\bra\phi^p_f \phi_f^{2q+1}\phi^p_f \phi_f^{2q+1}\ket$ and
    ${\sf c}_{0,q,p,\ell}\equiv{\sf c}_{q,p,\ell}$. Unfortunately we did
    not find a closed expression for these coefficients for general
    $q$ and $\ell$. 

    The matching conditions (\ref{eq:match}) can be written more precisely
    in the form
    \beq
    \lim_{\epsilon\to0}\frac{{\sf M}_{p,\ell}^2{\sf Q}}{\gamma_\phi}=
    \left(\frac{{\sf c}_{q,p,\ell}}{{\sf c}_{p,\ell}}\right)^2\,,
    \label{eq:Match}
    \eeq
    where ${\sf M}_{p,\ell}$ and ${\sf Q}$ are defined in (\ref{eq:M}) and
    (\ref{eq:Q}).

    When $\ell=0$, then $\gamma_{p,0}\equiv\gamma_{\phi^{p+1}}$ and Eq.s (\ref{eq:cpl}) and (\ref{eq:cq}) give at once
    \beq
    \frac{{\sf c}_{q,p,0}}{{\sf c}_{p,0}}=
    \left(\begin{matrix}p\cr
    q\cr\end{matrix}\right)\frac{\sqrt{(2q+1)!}}{(q+1)!}\,.
      \eeq
      In such a case the matching conditions define a set of recursion relations that have been completely solved in  \cite{Gliozzi:2017hni}. The solution reads
      \beq
      \gamma^{(1)}_{\phi^p}=\frac{q(p-q)_{q+1}}{(q+1)_{q+1}}~,~
      \gamma^{(2)}_\phi=2\,q^2\left[\frac{((q+1)!)^2}{(2(q+1))!}\right]^3\,.
      \label{eq:solution0}
      \eeq
      Thus, the only remaining unknown in (\ref{eq:Match}) is $\gamma_{p,\ell}$. This equation can be further simplified considering the four-point function $\bra\phi^q\,\phi^{q+1}\,\oo_{p,\ell}\,\phi^p\ket$. The associated matching condition reads
 \beq
 \lim_{\epsilon\to0}\frac{{\sf M}_{q,0}  {\sf M}_{p,\ell} Q}{\gamma_\phi}=
 \frac{{\sf c}_{q,p,\ell}}{{\sf c}_{p,\ell}}
 \frac{{\sf c}_{q,q,0}}{{\sf c}_{q,0}}\,.
 \eeq
 Now this equation is linear in $\gamma_{p,\ell}$ and the solution is
\beq
\gamma_{p,\ell}^{(1)}=\gamma_{\phi^p}^{(1)}+4\gamma^{(2)}_\phi\frac{{\sf c}_{q,p,\ell}}
{{\sf c}_{p,\ell}}
\frac{1+q}{q(1+q\,\ell)}\frac{\left(\sqrt{(2q+1)!}\right)^3}
     {((q+1)!)^2}\,,
\label{eq:solutionL}     
\eeq
were $q$ and $d$ are related by $d=2(q+1)/q$; ${\sf c}_{p,\ell}$ is given in (\ref{eq:cpl}) and  ${\sf c}_{q,p,\ell}$ is implicitly defined in (\ref{eq:cq}).
Actually,  when $\oo_{p,\ell}$ is degenerate this equation is of no practical use. In such a case the
fusion rule (\ref{eq:fusionspin}) we used to construct $\oo_{p,\ell}$  does not
select a single primary, but a linear combination of them; these have different anomalous dimensions in $d-\epsilon$, therefore 
 when both $p$ and $\ell$ are large it does not apply\footnote{I thank L.F. Alday for this
  important remark and for pointing out Ref.\cite{Kehrein:1994ff} to me. Note
   that eq.(\ref{eq:solutionL}) is still true even in the degenerate
  case, provided we assume that $\oo_{p,\ell}$ is an eigenfunction of $D$  in $d-\epsilon$ and the OPE coefficients  ${\sf c}_{p,\ell}$ and
  ${\sf c}_{q,p,\ell}$ are computed with this operator.}. Let us see some specific
examples where it works. There is no primary of spin $\ell=1$ in a theory with a single real scalar. For  $d=4-\epsilon$  we have for any $p$ and any $\ell>1$
\beq
\gamma_{p,2}^{(1)}=\frac{(p-1)(4+3p)}{18}\,,~\gamma_{p,3}^{(1)}=\frac{(p^2-3)}{6}\,,~\gamma_{2,\ell}^{(1)}=\frac13+\frac{2\,(-1)^\ell}{3\,(\ell+1)}.
\label{eq:spind4}
\eeq 
Similarly for $d=3-\epsilon$
\beq
\gamma_{p,2}^{(1)}=\frac{(p-1)(p-2)(5p+18)}{150}\,,~\gamma_{p,3}^{(1)}=
\frac{(p-2)(7p^2+11p-90)}{210}\,,~\gamma_{2,\ell}^{(1)}=0\,.
\label{eq:spind3}
\eeq
In $d=4-\epsilon$ and $p<12$ these $\gamma$'s have been also computed in
\cite{Kehrein:1994ff} and the results agree.
 When $\ell>3$ and $p>2$ these operators are degenerate in the free theory limit. For instance for $\ell=4$ and $p>2$ the space of primaries has
dimension two \cite{Kehrein:1994ff}, while the equation above reads   
\beq
\gamma_{p,4}^{(1)}=\frac{p(p-1)}{6}+\frac{\sqrt{2(p-1)(2p^3-11p^2+78p-104)}}
{15\sqrt{p^2+p+6}}\,.
\eeq
It gives the exact value for $p=2$ and an intermediate value between the two
exact $\gamma$'s in the other cases.
For large values of $\ell$ one could apply the inversion formula of
\cite{Caron-Huot:2017vep} (see also \cite{Simmons-Duffin:2017nub})
 to extract the OPE coefficients of (\ref{eq:cq}) in this limit.

In accordance with the theorem (\ref{eq:theorem}) we have, in the case $p=1$,
\beq
 {\sf c}_{q,1,\ell}\equiv {\sf c}_{\cJ_\ell\,\phi\,\phi^{2q+1}}=0\,,
\eeq
moreover $\gamma^{(1)}_\phi=0$, then $\gamma_{p,\ell}^{(1)}=0$
\footnote{This is only true for the upper critical dimensions $d=2(q+1)/q$. For the more general case $d=2m/(m-2)$, which includes $d=6$, the above matching conditions do not apply. Nevertheless we shall extract the anomalous dimensions 
of $\cJ_\ell$ also in such a case.}.

 Further information on the anomalous dimensions of $\cJ_\ell$ can be gained 
by studying the five point function $\bra\phi^q\,\phi^{q+1}\cJ_\ell\,\phi^q\,
\phi^{q+1}\ket$. In the factorization scheme shown in Figure 1 we have
\beq
\bra\phi^q\phi^{q+1}\cJ_\ell\,\phi^q
\phi^{q+1}\ket=\sum_{\oo}\sum_{\oo'}{\sf c}_\oo{\sf c}_{\oo'}{\sf c}_\cJ\sum_{\alpha\in\cH_\oo}\sum_{\beta\in\cH_{\oo'}}
\frac{\bra\phi^q\,\phi^{q+1}\vert
\alpha\ket\bra\alpha\vert\cJ_\ell\vert\beta\ket\bra
\beta\vert\phi^q\,\phi^{q+1}\ket}{\bra\alpha\vert\alpha\ket\,\bra\beta\vert\beta\ket}\,,
\eeq
where ${\sf c}_{\oo}\equiv{\sf c}_{\oo\phi^q\phi^{q+1}}$ and 
${\sf c}_{\cJ}\equiv{\sf c}_{\oo\oo'\cJ_\ell}$. In view of the already mentioned
theorem (\ref{eq:theorem}), if 
$\cJ_\ell$ is conserved and $\oo$, $\oo'$ are scalars, then  
${\sf c}_{\oo\oo'\cJ_\ell}\propto\delta_{\oo,\oo'}$. 
In particular, in the free field 
theory the two factorization channels always contain  the scalars $\phi_f$ and
$\phi^{2q+1}_f$. The OPE coefficients of these two contributions are
respectively
${\sf c}_{\phi_f\,\phi^q_f\,\phi^{q+1}_f}^2{\sf c}_{\phi_f\,\phi_f\,\cJ_\ell}\equiv
{\sf c}_{q,0}^2{\sf c}_{\phi_f\,\phi_f\,\cJ_\ell}$
and ${\sf c}_{\phi_f^{2q+1}\,\phi^q_f\,\phi^{q+1}_f}^2{\sf c}_{\phi_f^{2q+1}\,\phi_f^{2q+1}\,
  \cJ_\ell}=
(2q+1){\sf c}_{q,q,0}^2{\sf c}_{\phi_f\,\phi_f\,\cJ_\ell}$,   where we used the
useful identity (see  Appendix)
\beq
\frac{{\sf c}_{\phi^{p}_f\phi^{p}_f\cJ_\ell}}{{\sf c}_{\phi_f\,\phi_f\,\cJ_\ell}}=p\,.
\label{eq:ida}
\eeq 
 \FIGURE{
\setlength{\unitlength}{1cm}
\thicklines
\begin{picture}(5,6)
  \put(4.5,3){$\phi^{q}$}
  \put(4.5,1.1){$\phi^{q+1}$}
    \put(.5,3){$\phi^{q+1}$}
    \put(.5,1.1){$\phi^{q}$}
   \put(2.6,3){$\cJ_\ell$}  
%{\color[rgb]{.56,0,.t6}  
\put(1,2.2){\line(1,0){3}}
\put(4,2.2){\line(2,3){.5}}
\put(4,2.2){\line(2,-3){.5}}
\put(2.5,2.2){\line(0,1){1}}
\put(1,2.2){\line(-2,3){.5}}
\put(1,2.2){\line(-2,-3){.5}}
%}
\end{picture}
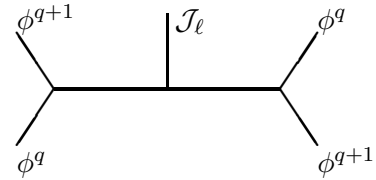
\caption{The factorization channels of the five-point  under study.}}
   
Turning on the interaction, $\phi^{2q+1}$ becomes a descendant of $\phi$ and the second contribution has to be absorbed by the first, in accordance  with (\ref{eq:c2toc1}). Therefore the matching condition of the five-point function under study  
takes the form
\beq
\lim_{\epsilon\to0}\frac{{M}_{q,0}^2\,{\sf N}_\ell\,{\sf Q}^2}{\gamma_\phi^2}=(2q+1)
\left(\frac{{\sf c}_{q,q,0}}{{\sf c}_{q,0}}\right)^2\,.
\label{eq:Matching5}
\eeq
When combined with (\ref{eq:Match}) it leads to
\beq
\lim_{\epsilon\to0}\frac{{\sf N}_\ell\,{\sf Q}}{\gamma_\phi}=2q+1\,.
\label{eq:Natching5}
\eeq
On substituting (\ref{eq:N}) we find at once
\beq
\gamma^{(2)}_\ell=2\,\gamma^{(2)}_\phi\left(1-\frac{(\nu+1)(\nu+2)}
{(\ell+\nu)(\ell+\nu-1)}\right)\,,
\label{eq:HS}
\eeq
with $\nu=\frac d2-1=\frac 1q$. The above reproduces the correct second-order anomalous dimension of $\cJ_\ell$ at
$d=4-\epsilon$ \cite{Wilson:1973jj}. This expression is true  more generally at the 
multicritical WF fixed points at $d=2(q+1)/q-\epsilon$  which include the
physically relevant case at $d=3-\epsilon$. In all the cases we have
$\gamma^{(2)}_2=0$, as expected by the conservation of the stress tensor at the WF fixed point.

There is a simple modification that allows to generalize the above matching condition to the general case of the upper critical dimension $d_m=2m/(m-2)$. Let us see for simplicity the case $d=6-\epsilon$, where the scalar becoming a descendant of $\phi$ is $\phi^2$. At the Gaussian fixed point we have the fusion rule 
$[\phi_f]\times[\phi_f]=1+\sqrt{2}\,[\phi_f^2]+{\rm spinning\,\, primaries}$. In the interacting theory $\phi^2$ is no longer a primary, hence it should 
be replaced by $\phi$, but   $\lim_{\epsilon\to0}{\sf c}_{\phi\phi\phi}=0$. The
resulting matching condition of the four-point $\bra \phi\phi\phi\phi\ket$ reads
\footnote{Such a relation has been first pointed out by Yu Nakayama in some
  unpublished notes. I thank him for sharing with me his notes.}  
\beq
\lim_{\epsilon\to 0}{\sf c}_{\phi\phi\phi}^2\frac{{\sf Q}\,\D_\phi(\D_\phi+2-d)}
    {\gamma_\phi}=2\,.
    \eeq
    It can be used to express the OPE coefficient ${\sf c}_{\phi\phi\phi}$ of the interacting theory in terms of the anomalous dimensions of $\phi$. 
More generally, at a WF fixed point in which the primary descendant
$\phi^{m-1}$ is a even power $m-1=2q$ we have
\beq
\lim_{\epsilon\to 0}{\sf c}_{\phi^q\phi^q\phi}^2\frac{{\sf Q}\,\D_\phi(\D_\phi+2-d)}
    {\gamma_\phi}=\frac{(2q)!}{(q!)^2}\,,
    \label{eq:even}
    \eeq
    with $d=\frac{2(2q+1)}{2q-1}-\epsilon$.
    The matching condition of the five-point function
    $\bra\phi^q\phi^q \cJ_\ell\,\phi^q\phi^q\ket$, when combined with
    (\ref{eq:even}) and (\ref{eq:ida}) gives the promised generalization
    \beq
\lim_{\epsilon\to0}\frac{{\sf N}_\ell\,{\sf Q}}{\gamma_\phi}=m-1=\frac{d_m+2}{d_m-2}\,,
\label{eq:N5}
\eeq
valid for any integer $m$. Notice however that the first
non-vanishing anomalous dimensions of $\cJ_\ell$ in $d=6-\epsilon$ are
$O(\epsilon)$, since in this case, unlike what happens at
$d=4-\epsilon$, at $d=3-\epsilon$ and at the other multicritical
fixed points described by (\ref{eq:solution0}), now
$\gamma^{(1)}\not=0$. Precisely we have
$\gamma^{(1)}_\phi=-\frac1{18}$, so the  equation above yields
\beq
\gamma_\ell^{(1)}=-\frac19\left(1-\frac{12}{(\ell+1)(\ell+2)}\right)\,,
\eeq
which exactly agrees with the known results \cite{Giombi:2016hkj}.

The matching condition (\ref{eq:N5}) applies also when $\ell=0$, where it gives
a new equation\footnote{The only exception is $d=6-\epsilon$ because $\phi^2$ is not a primary there.} for the anomalous dimension of $\cJ_0\equiv\phi^2$. This is useful in the multicritical case, where $\gamma^{(1)}_{\phi^2}$ is vanishing (see eq. (\ref{eq:solution0})). This new equation gives
\beq
\gamma_{\phi^2}^{(2)}=\frac{4 q (2  + q)}{(q-1)}\gamma_{\phi}^{(2)}=
\frac{8 q^3(q+2)}{q-1}\left[\frac{((q+1)!)^2}{(2(q+1))!}\right]^3\,.
\label{eq:multi}
\eeq
In particular the $d=3$ scalar theory perturbed with a $\phi^6$ potential
(i.e. $q=2$) gives
$\gamma_{\phi^2}^{(2)}=\frac4{125}\,$, as can be directly checked by putting $\ell=0$ in (\ref{eq:HS}). Eq.(\ref{eq:multi}) agrees with the results obtained with more conventional methods based on the Lagrangian formulation and
 equations of motion
\cite{Codello:2017qek}.

\section{Conclusions}
In the present paper we  studied some physical properties of perturbative
Wilson-Fisher fixed points under the only assumption of their conformal invariance. Using a generalization of the method developed in
\cite{Gliozzi:2016ysv,Gliozzi:2017hni}, we computed, to the first non-trivial order in the $\epsilon$ expansion, the anomalous dimensions of a large set 
spinning operators $\oo_{p,\ell}$ that in the Gaussian theory have scaling dimensions
$\D=(p+1)(\frac d2-1)+\ell$. In accordance with the bootstrap philosophy no Lagrangian was assumed. The interaction was implicitly turned on by assuming that the local operators acquire anomalous dimensions. The WF fixed points were defined using only conformal invariant concepts, with no reference to the
renormalization group. They were seen as
smooth conformal deformations of scalar free-field theories (see the precise
definition around Eq. (\ref{eq:anomalous})),
where it was somewhat hidden the notion of UV and IR fixed points of the RG.

All the computations of the present paper rely on a unique group-theoretical mechanism, the conformal multiplet recombination. Schematically, it tells us that
the free field scalar $\phi_f$ corresponds to a short multiplet, while
the multiplet of the corresponding interacting field is long; in the $\epsilon\to0$ limit thy must agree. We derived in this way two sets
of matching conditions --(\ref{eq:Match}) and  (\ref{eq:N5}) --
which encode the wanted information on $\oo_{\D,\ell}$.
The former describes the matching conditions of certain four-point functions
which generalize to $\ell\not=0$ those already considered in
\cite{Gliozzi:2016ysv,Gliozzi:2017hni}.  The latter is completely new and is
obtained by  matching conditions of the five-point function
sketched in  figure 1. We used it to compute the anomalous dimensions
of the subclass of spinning operators $\cJ_\ell\equiv\oo_{1,\ell}$; they are
conserved an the Gaussian fixed point, while acquire anomalous dimensions at the WF fixed point. The resulting single equation (\ref{eq:N5}) tells us that
the stress tensor is
automatically conserved at any (permitted) space-time dimension,
while for $\ell\not=2$ yields the  first non-vanishing contribution
of the anomalous dimensions of the HS currents. In the more conventional
approaches these results require different Lagrangians with different perturbing
potentials, namely $\phi^4$ in
 $d=4-\epsilon$, $\phi^6$ in $d=3-\epsilon$, and  $\phi^3$ in 
$d=6-\epsilon$. 

There are several directions one can pursue. The most obvious one is to
extend our analysis to $O(N)$ invariant models. Another interesting
extension would be to apply this approach to generalized free-field theories,
following the line
initiated in  \cite{Gliozzi:2016ysv,Gliozzi:2017hni} for the scalar primaries.
Notice that in the present approach, unlike other conformal bootstrap
calculations, the crossing symmetry has not been used. Enforcing 
this powerful symmetry in our scheme could hopefully generate new results.

It would be finally important to find some connection of our method
to other bootstrap-inspired analytic approaches. In an elegant study which relies on turning the crossing equations into an algebraic problem \cite{Alday:2016njk,Alday:2016jfr}, the HS symmetry is explicitly broken by introducing a small parameter $g$.
At first non-trivial  order in $g$ the spectrum of weakly broken currents 
follows. Unfortunately this method cannot apply to more general, non conserved, 
spinning operators; notice also that the conservation of the stress tensor
does not follow automatically, and has to
be assumed as a further constraint. It would be  interesting to
combine such an algebraic form of the crossing equations with the
mechanism of conformal multiplet recombination used here. Another powerful way to obtain analytic information about the spectrum of the primary operators in a WF fixed point relies
upon a Mellin space representation of correlation functions 
\cite{Gopakumar:2016wkt,Gopakumar:2016cpb,Dey:2016mcs}. It gives at the
moment the analytic results of higher order in the anomalous dimensions and
OPE coefficients. So far it has been only applied to the four-point function
$\bra\phi\phi\phi\phi\ket$. An extension of the analysis to
$\bra\phi^p\phi\,\phi^p\phi\ket$  would provide information on a much larger
class of spinning operators.
\section*{Acknowledgements}
The author is grateful to A. L. Guerrieri, A. C. Petkou and C. Wen for fruitful  discussions and comments on the manuscript. He would also like  to thank
L. F. Alday, S. Hikami and M. Meineri for useful discussions and correspondence.
\appendix
\section{OPE coefficients of the free-field theory}
In this appendix we collect some useful formulae on the conformal block expansions and OPE coefficients of a scalar free field theory. Here $\phi$ is the free fundamental scalar and $\phi^p$ denotes the normal ordered scalar primary
$\frac{:(\phi)^p:}{\sqrt{p!}}$. The normalization is chosen in such a way that
\beq
\bra\phi^p(x_1)\phi^p(x_2)\ket=\frac1{(x_{12}^2)^{p\,\nu}}\,.
\eeq
Applying the Wick's contractions to the four-point function
$\bra\phi^p(x_1)\phi^r(x_2)\phi^p(x_3)\phi^r(x_4)\ket$ with $p\ge r$ and
comparing it
with the standard parameterization of a generic CFT, namely,
\beq
\bra \oo_1(x_1)\oo_2(x_2)\oo_3(x_3)\oo_4(x_4)\ket=\frac{g(u,v)}
     {\vert x_{12}\vert^{\D^{+}_{12}}\vert x_{34}\vert^{\D^{+}_{34}}}
     \left( {\vert x_{24}\vert \over \vert x_{14}\vert} \right)^{\D^{-}_{12}}
\left( {\vert x_{14}\vert \over \vert x_{13}\vert} \right)^{\D^{-}_{34}} 
\,,
\label{eq:4pt}
\eeq
where $\D^{\pm}_{ij} = \Delta_{i} \pm \D_{j}$, $i,j=1,2,3,4$ and $\D_{i}$ is the
 scaling dimension of $\oo_i$, while $u={x_{12}^2 x_{34}^2 \over x_{13}^2 x_{
     24}^2}$ and
 $v={x_{14}^2 x_{23}^2 \over x_{13}^2 x_{24}^2}$ are the cross ratios,
 we have
 \beq
 g_{p,r}(u,v)=\sum_{h=0}^r{\sf g}_h(u,v)\equiv\sum_{h=0}^r\frac{p!r!}{(h!)^2}u^{(p+r-2h)\nu/2}\sum_{k=0}^{r-h}
 \frac{v^{-k\nu}}{(k!)^2(p-h-k)!(r-h-k)!}\,.
 \label{eq:g}
 \eeq
 Each ${\sf g}_h(u,v)$ can be expanded in terms of conformal blocks
 $G^{a,b}_{\D,\ell}(u,v)$, i.e. eigenfunctions of the quadratic and quartic Casimir invariants of the conformal group. $\D$ and $\ell$ are the scaling dimension and the spin of the primaries that could appear in the OPE of
 $\phi^p(x_1)\,\phi^r(x_2)$ and  $a=\nu\frac{r-p}2,\,  b=-a$. The set of primaries contributing to  ${\sf g}_h(u,v)$ belongs to the linear Regge trajectory 
$\D_{n,\ell}=(p+r-2h+2n)\nu+\ell$:
 \beq
 {\sf g}_h(u,v)=\sum_{n,\ell}^\infty{\sf c}^2_{\phi^p\,\phi^r\oo_{\D_{n,\ell}}}
 G^{a,b}_{\D_{n,\ell}}(u,v)\,.
 \label{eq:gexpansion}
 \eeq
 Knowing the spectrum of the contributing primaries one may compute term by term the associated OPE coefficients, using for instance the method described in appendix $A$ of \cite{Gliozzi:2017hni}. In some cases from the first few
 OPE coefficients one can infer the general form for any $\ell$ and $n$.
 In particular if we put $r=1$ in (\ref{eq:g}) we find
 \beq
 g_{p,1}(u,v)=p\,u^{\nu(p-1)/2}+u^{\nu(p+1)/2}\left(1+\frac p{v^\nu}\right)\,.
 \eeq
 The first term describes a single primary, as
\beq
 u^{\nu(p-1)/2}=G^{a,-a}_{\nu(p-1),0}(u,v)\,,~~~~a=\frac{\nu(1-p)}2\,,
 \eeq
 while the primaries contributing to the second term belong to the linear Regge
 trajectory $\D_\ell\equiv\D_{0,\ell}=(p+1)\nu+\ell$. They coincide with
 the set of spinning operators considered in the present paper.
 It is not difficult to verify, using the method of \cite{Gliozzi:2017hni},
 that associated OPE coefficients are those given in (\ref{eq:cpl}).

 In the study of the five-point function we need to compute the OPE coefficient
 ${\sf c}_{\phi^p\phi^p\cJ_\ell}$, where the conserved HS current $\cJ_\ell$ has
 scaling dimension $2\nu+\ell$. This can be extracted from the conformal block
 expansion of $g_{p,p}(u,v)$, more precisely from the term ${\sf g}_{p-1}$ of
 (\ref{eq:g}). We have
 \beq
 {\sf g}_{p-1}(u,v)=p^2\,u^\nu\left(1+\frac1{v^\nu}\right)\,.
 \eeq
 Apart from the $p^2$ factor, it coincides with the $g$ function of
 $\bra\phi(x_1)\phi(x_2)\phi(x_3)\phi(x_4)\ket$, therefore we can infer at once eq.
 (\ref{eq:ida}).
\bibliographystyle{abe}

\begin{thebibliography}{20}


\bibitem{Ferrara:1973yt}
S.~Ferrara, A.~Grillo, and R.~Gatto, \emph{{Tensor representations of conformal
  algebra and conformally covariant operator product expansion}},
\href{http://dx.doi.org/10.1016/0003-4916(73)90446-6}{Annals Phys. {\bf 76}
  (1973)  161--188}.
%%CITATION = APNYA,76,161;%%.

\bibitem{Polyakov:1974gs}
A.~Polyakov, \emph{{Nonhamiltonian approach to conformal quantum field
  theory}},
Zh.Eksp.Teor.Fiz. {\bf 66} (1974)  23--42.
%%CITATION = ZETFA,66,23;%%.

\bibitem{Belavin:1984vu}
A.~Belavin, A.~M. Polyakov, and A.~Zamolodchikov, \emph{{Infinite Conformal
  Symmetry in Two-Dimensional Quantum Field Theory}},
\href{http://dx.doi.org/10.1016/0550-3213(84)90052-X}{Nucl.Phys. {\bf B241}
  (1984)  333--380}.
%%CITATION = NUPHA,B241,333;%%.

\bibitem{Rattazzi:2008pe}
R.~Rattazzi, V.~S. Rychkov, E.~Tonni, and A.~Vichi, \emph{{Bounding scalar
  operator dimensions in 4D CFT}},
  \href{http://dx.doi.org/10.1088/1126-6708/2008/12/031}{JHEP {\bf 0812} (2008)
   031},
\href{http://arxiv.org/abs/0807.0004}{{\tt arXiv:0807.0004 [hep-th]}}.
%%CITATION = ARXIV:0807.0004;%%.
\bibitem{Rychkov:2009ij}
V.~S. Rychkov and A.~Vichi, \emph{{Universal Constraints on Conformal Operator
  Dimensions}}, \href{http://dx.doi.org/10.1103/PhysRevD.80.045006}{Phys.Rev.
  {\bf D80} (2009)  045006},
\href{http://arxiv.org/abs/0905.2211}{{\tt arXiv:0905.2211 [hep-th]}}.
%%CITATION = ARXIV:0905.2211;%%.

\bibitem{Rattazzi:2010gj}
R.~Rattazzi, S.~Rychkov, and A.~Vichi, \emph{{Central Charge Bounds in 4D
  Conformal Field Theory}},
  \href{http://dx.doi.org/10.1103/PhysRevD.83.046011}{Phys.Rev. {\bf D83}
  (2011)  046011},
\href{http://arxiv.org/abs/1009.2725}{{\tt arXiv:1009.2725 [hep-th]}}.
%%CITATION = ARXIV:1009.2725;%%.

\bibitem{Poland:2010wg}
D.~Poland and D.~Simmons-Duffin, \emph{{Bounds on 4D Conformal and
  Superconformal Field Theories}},
  \href{http://dx.doi.org/10.1007/JHEP05(2011)017}{JHEP {\bf 1105} (2011)
  017},
\href{http://arxiv.org/abs/1009.2087}{{\tt arXiv:1009.2087 [hep-th]}}.
%%CITATION = ARXIV:1009.2087;%%.

\bibitem{ElShowk:2012ht}
S.~El-Showk, M.~F. Paulos, D.~Poland, S.~Rychkov, D.~Simmons-Duffin, {\em et
  al.}, \emph{{Solving the 3D Ising Model with the Conformal Bootstrap}},
  \href{http://dx.doi.org/10.1103/PhysRevD.86.025022}{Phys.Rev. {\bf D86}
  (2012)  025022},
\href{http://arxiv.org/abs/1203.6064}{{\tt arXiv:1203.6064 [hep-th]}}.
%%CITATION = ARXIV:1203.6064;%%.
\bibitem{Liendo:2012hy}
P.~Liendo, L.~Rastelli, and B.~C. van Rees, \emph{{The Bootstrap Program for
  Boundary $CFT_d$}}, \href{http://dx.doi.org/10.1007/JHEP07(2013)113}{JHEP
  {\bf 1307} (2013)  113},
\href{http://arxiv.org/abs/1210.4258}{{\tt arXiv:1210.4258 [hep-th]}}.
%%CITATION = ARXIV:1210.4258;%%.

\bibitem{Pappadopulo:2012jk}
D.~Pappadopulo, S.~Rychkov, J.~Espin, and R.~Rattazzi, \emph{{OPE Convergence
  in Conformal Field Theory}},
  \href{http://dx.doi.org/10.1103/PhysRevD.86.105043}{Phys.Rev. {\bf D86}
  (2012)  105043},
\href{http://arxiv.org/abs/1208.6449}{{\tt arXiv:1208.6449 [hep-th]}}.
%%CITATION = ARXIV:1208.6449;%%.

\bibitem{ElShowk:2012hu}
S.~El-Showk and M.~F. Paulos, \emph{{Bootstrapping Conformal Field Theories
  with the Extremal Functional Method}},
  \href{http://dx.doi.org/10.1103/PhysRevLett.111.241601}{Phys.Rev.Lett. {\bf
  111} (2013) no.~24, 241601},
\href{http://arxiv.org/abs/1211.2810}{{\tt arXiv:1211.2810 [hep-th]}}.
%%CITATION = ARXIV:1211.2810;%%.

\bibitem{Gliozzi:2013ysa}
F.~Gliozzi, \emph{{More constraining conformal bootstrap}},
  \href{http://dx.doi.org/10.1103/PhysRevLett.111.161602}{Phys.Rev.Lett. {\bf
  111} (2013)  161602},
\href{http://arxiv.org/abs/1307.3111}{{\tt arXiv:1307.3111}}.
\bibitem{Gaiotto:2013nva}
D.~Gaiotto, D.~Mazac, and M.~F. Paulos, \emph{{Bootstrapping the 3d Ising twist
  defect}}, \href{http://dx.doi.org/10.1007/JHEP03(2014)100}{JHEP {\bf 1403}
  (2014)  100},
\href{http://arxiv.org/abs/1310.5078}{{\tt arXiv:1310.5078 [hep-th]}}.
%%CITATION = ARXIV:1310.5078;%%.

\bibitem{El-Showk:2014dwa}
S.~El-Showk, M.~F. Paulos, D.~Poland, S.~Rychkov, D.~Simmons-Duffin, and
  A.~Vichi, \emph{{Solving the 3d Ising Model with the Conformal Bootstrap II.
  c-Minimization and Precise Critical Exponents}},
  \href{http://dx.doi.org/10.1007/s10955-014-1042-7}{J. Stat. Phys. {\bf 157}
  (2014)  869},
\href{http://arxiv.org/abs/1403.4545}{{\tt arXiv:1403.4545 [hep-th]}}.
%%CITATION = ARXIV:1403.4545;%%.

\bibitem{Beem:2013qxa}
C.~Beem, L.~Rastelli, and B.~C. van Rees, \emph{{The $\mathcal N=4$
  Superconformal Bootstrap}},
  \href{http://dx.doi.org/10.1103/PhysRevLett.111.071601}{Phys.Rev.Lett. {\bf
  111} (2013)  071601},
\href{http://arxiv.org/abs/1304.1803}{{\tt arXiv:1304.1803 [hep-th]}}.
%%CITATION = ARXIV:1304.1803;%%.

\bibitem{Nakayama:2014yia}
Y.~Nakayama and T.~Ohtsuki, \emph{{Five dimensional $O(N)$-symmetric CFTs from
  conformal bootstrap}},
  \href{http://dx.doi.org/10.1016/j.physletb.2014.05.058}{Phys.Lett. {\bf B734}
  (2014)  193--197},
\href{http://arxiv.org/abs/1404.5201}{{\tt arXiv:1404.5201 [hep-th]}}.
%%CITATION = ARXIV:1404.5201;%%.
\bibitem{Nakayama:2014sba}
Y.~Nakayama and T.~Ohtsuki, \emph{{Bootstrapping phase transitions in QCD and
  frustrated spin systems}},
  \href{http://dx.doi.org/10.1103/PhysRevD.91.021901}{Phys. Rev. {\bf D91}
  (2015) no.~2, 021901},
\href{http://arxiv.org/abs/1407.6195}{{\tt arXiv:1407.6195 [hep-th]}}.
%%CITATION = ARXIV:1407.6195;%%.

\bibitem{Gliozzi:2014jsa}
F.~Gliozzi and A.~Rago, \emph{{Critical exponents of the 3d Ising and related
  models from Conformal Bootstrap}},
  \href{http://dx.doi.org/10.1007/JHEP10(2014)042}{JHEP {\bf 1410} (2014)  42},
\href{http://arxiv.org/abs/1403.6003}{{\tt arXiv:1403.6003 [hep-th]}}.
%%CITATION = ARXIV:1403.6003;%%.

\bibitem{Chester:2014fya}
S.~M. Chester, J.~Lee, S.~S. Pufu, and R.~Yacoby, \emph{{The $ \mathcal{N}=8 $
  superconformal bootstrap in three dimensions}},
  \href{http://dx.doi.org/10.1007/JHEP09(2014)143}{JHEP {\bf 1409} (2014)
  143},
\href{http://arxiv.org/abs/1406.4814}{{\tt arXiv:1406.4814 [hep-th]}}.
%%CITATION = ARXIV:1406.4814;%%.

\bibitem{Kos:2014bka}
F.~Kos, D.~Poland, and D.~Simmons-Duffin, \emph{{Bootstrapping Mixed
  Correlators in the 3D Ising Model}},
  \href{http://dx.doi.org/10.1007/JHEP11(2014)109}{JHEP {\bf 11} (2014)  109},
\href{http://arxiv.org/abs/1406.4858}{{\tt arXiv:1406.4858 [hep-th]}}.
%%CITATION = ARXIV:1406.4858;%%.
\bibitem{Chester:2014gqa}
S.~M. Chester, S.~S. Pufu, and R.~Yacoby, \emph{{Bootstrapping O(N) Vector
    Models in $4 < d < 6$}},
\href{http://dx.doi:10.1103/PhysRevD.91.086014}{Phys.\ Rev.\ D {\bf 91}
  (2015) no.8,  086014},
\href{http://arxiv.org/abs/1412.7746}{{\tt arXiv:1412.7746 [hep-th]}}.
%%CITATION = ARXIV:1412.7746;%%.

\bibitem{Beem:2014zpa}
C.~Beem, M.~Lemos, P.~Liendo, L.~Rastelli, and B.~C. van Rees, \emph{{The $
  \mathcal{N}=2 $ superconformal bootstrap}},
  \href{http://dx.doi.org/10.1007/JHEP03(2016)183}{JHEP {\bf 03} (2016)  183},
\href{http://arxiv.org/abs/1412.7541}{{\tt arXiv:1412.7541 [hep-th]}}.
%%CITATION = ARXIV:1412.7541;%%.

\bibitem{Simmons-Duffin:2015qma}
D.~Simmons-Duffin, \emph{{A Semidefinite Program Solver for the Conformal
  Bootstrap}}, \href{http://dx.doi.org/10.1007/JHEP06(2015)174}{JHEP {\bf 06}
  (2015)  174},
\href{http://arxiv.org/abs/1502.02033}{{\tt arXiv:1502.02033 [hep-th]}}.
%%CITATION = ARXIV:1502.02033;%%.

\bibitem{Bobev:2015vsa}
N.~Bobev, S.~El-Showk, D.~Mazac, and M.~F. Paulos, \emph{{Bootstrapping the
  Three-Dimensional Supersymmetric Ising Model}},
  \href{http://dx.doi.org/10.1103/PhysRevLett.115.051601}{Phys. Rev. Lett. {\bf
  115} (2015) no.~5, 051601},
\href{http://arxiv.org/abs/1502.04124}{{\tt arXiv:1502.04124 [hep-th]}}.
%%CITATION = ARXIV:1502.04124;%%.

\bibitem{Kos:2015mba}
F.~Kos, D.~Poland, D.~Simmons-Duffin, and A.~Vichi, \emph{{Bootstrapping the
  O(N) Archipelago}}, \href{http://dx.doi.org/10.1007/JHEP11(2015)106}{JHEP
  {\bf 11} (2015)  106},
\href{http://arxiv.org/abs/1504.07997}{{\tt arXiv:1504.07997 [hep-th]}}.
%%CITATION = ARXIV:1504.07997;%%.

\bibitem{Bobev:2015jxa}
N.~Bobev, S.~El-Showk, D.~Mazac, and M.~F. Paulos, \emph{{Bootstrapping SCFTs
  with Four Supercharges}},
  \href{http://dx.doi.org/10.1007/JHEP08(2015)142}{JHEP {\bf 08} (2015)  142},
\href{http://arxiv.org/abs/1503.02081}{{\tt arXiv:1503.02081 [hep-th]}}.
%%CITATION = ARXIV:1503.02081;%%.

\bibitem{Gliozzi:2015qsa}
F.~Gliozzi, P.~Liendo, M.~Meineri, and A.~Rago, \emph{{Boundary and Interface
  CFTs from the Conformal Bootstrap}},
  \href{http://dx.doi.org/10.1007/JHEP05(2015)036}{JHEP {\bf 05} (2015)  036},
\href{http://arxiv.org/abs/1502.07217}{{\tt arXiv:1502.07217 [hep-th]}}.
%%CITATION = ARXIV:1502.07217;%%.

\bibitem{Beem:2015aoa}
C.~Beem, M.~Lemos, L.~Rastelli, and B.~C. van Rees, \emph{{The (2, 0)
  superconformal bootstrap}},
  \href{http://dx.doi.org/10.1103/PhysRevD.93.025016}{Phys. Rev. {\bf D93}
  (2016) no.~2, 025016},
\href{http://arxiv.org/abs/1507.05637}{{\tt arXiv:1507.05637 [hep-th]}}.
%%CITATION = ARXIV:1507.05637;%%.

\bibitem{Nakayama:2016jhq}
Y.~Nakayama and T.~Ohtsuki, \emph{{Conformal Bootstrap Dashing Hopes of
    Emergent Symmetry}}, \href{http://dx.doi:10.1103/PhysRevLett.117.131601}{Phys.\ Rev.\ Lett.\  {\bf 117} (2016) no.13,  131601}
\href{http://arxiv.org/abs/1602.07295}{{\tt arXiv:1602.07295
  [cond-mat.str-el]}}.
%%CITATION = ARXIV:1602.07295;%%.

\bibitem{Kos:2016ysd}
F.~Kos, D.~Poland, D.~Simmons-Duffin, and A.~Vichi, \emph{{Precision islands in
  the Ising and O(N ) models}},
  \href{http://dx.doi.org/10.1007/JHEP08(2016)036}{JHEP {\bf 08} (2016)  036},
\href{http://arxiv.org/abs/1603.04436}{{\tt arXiv:1603.04436 [hep-th]}}.
%%CITATION = ARXIV:1603.04436;%%.
\bibitem{Nakayama:2016cim}
Y.~Nakayama, \emph{{Bootstrapping critical Ising model on three-dimensional
  real projective space}},
  \href{http://dx.doi.org/10.1103/PhysRevLett.116.141602}{Phys. Rev. Lett. {\bf
  116} (2016)  141602},
\href{http://arxiv.org/abs/1601.06851}{{\tt arXiv:1601.06851 [hep-th]}}.
%%CITATION = ARXIV:1601.06851;%%.

\bibitem{Gliozzi:2016cmg}
F.~Gliozzi, \emph{{Truncatable bootstrap equations in algebraic form and
  critical surface exponents}},
  \href{http://dx.doi.org/10.1007/JHEP10(2016)037}{JHEP {\bf 10} (2016)  037},
\href{http://arxiv.org/abs/1605.04175}{{\tt arXiv:1605.04175 [hep-th]}}.
%%CITATION = ARXIV:1605.04175;%%.

\bibitem{Esterlis:2016psv}
I.~Esterlis, A.~L. Fitzpatrick, and D.~Ramirez, \emph{{Closure of the Operator
  Product Expansion in the Non-Unitary Bootstrap}},
  \href{http://dx.doi.org/10.1007/JHEP11(2016)030}{JHEP {\bf 11} (2016)  030},
\href{http://arxiv.org/abs/1606.07458}{{\tt arXiv:1606.07458 [hep-th]}}.
%%CITATION = ARXIV:1606.07458;%%.

\bibitem{El-Showk:2016mxr}
S.~El-Showk and M.~F. Paulos, \emph{{Extremal bootstrapping: go with the
  flow}},
\href{http://arxiv.org/abs/1605.08087}{{\tt arXiv:1605.08087 [hep-th]}}.
%%CITATION = ARXIV:1605.08087;%%.
\bibitem{Hikami:2017hwv}
S.~Hikami, \emph{{Conformal Bootstrap Analysis for Yang-Lee Edge Singularity}},
\href{http://arxiv.org/abs/1707.04813}{{\tt arXiv:1707.04813 [hep-th]}}.
%%CITATION = ARXIV:1707.04813;%%.


\bibitem{Li:2017agi}
W.~Li, \emph{{Inverse Bootstrapping Conformal Field Theories}},
\href{http://arxiv.org/abs/1706.04054}{{\tt arXiv:1706.04054 [hep-th]}}.
%%CITATION = ARXIV:1706.04054;%%.

\bibitem{Wilson:1973jj}
K.~G. Wilson and J.~B. Kogut, \emph{{The Renormalization group and the epsilon
  expansion}},
\href{http://dx.doi.org/10.1016/0370-1573(74)90023-4}{Phys. Rept. {\bf 12}
  (1974)  75--200}.
%%CITATION = PRPLC,12,75;%%.

\bibitem{Rychkov:2015naa}
S.~Rychkov and Z.~M. Tan, \emph{{The $\epsilon$-expansion from conformal field
  theory}}, \href{http://dx.doi.org/10.1088/1751-8113/48/29/29FT01}{J. Phys.
  {\bf A48} (2015) no.~29, 29FT01},
\href{http://arxiv.org/abs/1505.00963}{{\tt arXiv:1505.00963 [hep-th]}}.
%%CITATION = ARXIV:1505.00963;%%.
\bibitem{Basu:2015gpa}
P.~Basu and C.~Krishnan, \emph{{$\epsilon$-expansions near three dimensions
  from conformal field theory}},
  \href{http://dx.doi.org/10.1007/JHEP11(2015)040}{JHEP {\bf 11} (2015)  040},
\href{http://arxiv.org/abs/1506.06616}{{\tt arXiv:1506.06616 [hep-th]}}.
%%CITATION = ARXIV:1506.06616;%%.

\bibitem{Nii:2016lpa}
K.~Nii, \emph{{Classical equation of motion and Anomalous dimensions at leading
  order}}, \href{http://dx.doi.org/10.1007/JHEP07(2016)107}{JHEP {\bf 07}
  (2016)  107},
\href{http://arxiv.org/abs/1605.08868}{{\tt arXiv:1605.08868 [hep-th]}}.
%%CITATION = ARXIV:1605.08868;%%.

\bibitem{Hasegawa:2016piv}
C.~Hasegawa and Y.~Nakayama, \emph{{$\epsilon$-Expansion in Critical
  $\phi^3$-Theory on Real Projective Space from Conformal Field Theory}},
\href{http://arxiv.org/abs/1611.06373}{{\tt arXiv:1611.06373 [hep-th]}}.
%%CITATION = ARXIV:1611.06373;%%.

\bibitem{Bashmakov:2016uqk}
V.~Bashmakov, M.~Bertolini, and H.~Raj, \emph{{Broken current anomalous
  dimensions, conformal manifolds, and renormalization group flows}},
  \href{http://dx.doi.org/10.1103/PhysRevD.95.066011}{Phys. Rev. {\bf D95}
  (2017) no.~6, 066011},
\href{http://arxiv.org/abs/1609.09820}{{\tt arXiv:1609.09820 [hep-th]}}.
%%CITATION = ARXIV:1609.09820;%%.
\bibitem{Gliozzi:2016ysv}
F.~Gliozzi, A.~Guerrieri, A.~C. Petkou, and C.~Wen, \emph{{Generalized
  Wilson-Fisher Critical Points from the Conformal Operator Product
  Expansion}}, \href{http://dx.doi.org/10.1103/PhysRevLett.118.061601}{Phys.
  Rev. Lett. {\bf 118} (2017) no.~6, 061601},
\href{http://arxiv.org/abs/1611.10344}{{\tt arXiv:1611.10344 [hep-th]}}.
%%CITATION = ARXIV:1611.10344;%%.

\bibitem{Liendo:2017wsn}
P.~Liendo, \emph{{Revisiting the dilatation operator of the Wilson–Fisher
  fixed point}}, \href{http://dx.doi.org/10.1016/j.nuclphysb.2017.04.020}{Nucl.
  Phys. {\bf B920} (2017)  368--384},
\href{http://arxiv.org/abs/1701.04830}{{\tt arXiv:1701.04830 [hep-th]}}.
%%CITATION = ARXIV:1701.04830;%%.

\bibitem{Gliozzi:2017hni}
F.~Gliozzi, A.~L. Guerrieri, A.~C. Petkou, and C.~Wen, \emph{{The analytic
  structure of conformal blocks and the generalized Wilson-Fisher fixed
  points}}, \href{http://dx.doi.org/10.1007/JHEP04(2017)056}{JHEP {\bf 04}
  (2017)  056},
\href{http://arxiv.org/abs/1702.03938}{{\tt arXiv:1702.03938 [hep-th]}}.
%%CITATION = ARXIV:1702.03938;%%.

\bibitem{Codello:2017qek}
A.~Codello, M.~Safari, G.~P. Vacca, and O.~Zanusso, \emph{{Leading CFT
  constraints on multi-critical models in d > 2}},
  \href{http://dx.doi.org/10.1007/JHEP04(2017)127}{JHEP {\bf 04} (2017)  127},
\href{http://arxiv.org/abs/1703.04830}{{\tt arXiv:1703.04830 [hep-th]}}.
%%CITATION = ARXIV:1703.04830;%%.

  \bibitem{Gracey:2017okb}
J.~A. Gracey, \emph{{Renormalization of scalar field theories in rational
  spacetime dimensions}},
\href{http://arxiv.org/abs/1703.09685}{{\tt arXiv:1703.09685 [hep-th]}}.
%%CITATION = ARXIV:1703.09685;%%.

\bibitem{Gopakumar:2016wkt}
R.~Gopakumar, A.~Kaviraj, K.~Sen, and A.~Sinha, \emph{{Conformal Bootstrap in
  Mellin Space}}, \href{http://dx.doi.org/10.1103/PhysRevLett.118.081601}{Phys.
  Rev. Lett. {\bf 118} (2017) no.~8, 081601},
\href{http://arxiv.org/abs/1609.00572}{{\tt arXiv:1609.00572 [hep-th]}}.
%%CITATION = ARXIV:1609.00572;%%.

\bibitem{Gopakumar:2016cpb}
R.~Gopakumar, A.~Kaviraj, K.~Sen, and A.~Sinha, \emph{{A Mellin space approach
  to the conformal bootstrap}},
  \href{http://dx.doi.org/10.1007/JHEP05(2017)027}{JHEP {\bf 05} (2017)  027},
\href{http://arxiv.org/abs/1611.08407}{{\tt arXiv:1611.08407 [hep-th]}}.
%%CITATION = ARXIV:1611.08407;%%.

\bibitem{Dey:2016mcs}
P.~Dey, A.~Kaviraj, and A.~Sinha, \emph{{Mellin space bootstrap for global
  symmetry}}, \href{http://dx.doi.org/10.1007/JHEP07(2017)019}{JHEP {\bf 07}
  (2017)  019},
\href{http://arxiv.org/abs/1612.05032}{{\tt arXiv:1612.05032 [hep-th]}}.
%%CITATION = ARXIV:1612.05032;%%.
\bibitem{Dey:2017oim}
P.~Dey and A.~Kaviraj, \emph{{Towards a Bootstrap approach to higher orders of
  epsilon expansion}},
\href{http://arxiv.org/abs/1711.01173}{{\tt arXiv:1711.01173 [hep-th]}}.
%%CITATION = ARXIV:1711.01173;%%.
\bibitem{Skvortsov:2015pea}
  E.~D.~Skvortsov,
  \emph{On (Un)Broken Higher-Spin Symmetry in Vector Models},
 \href{http://dx.doi:10.1142/9789813144101_0008}
  {{\tt arXiv:1512.05994 [hep-th]}}.
  %%CITATION = doi:10.1142/9789813144101_0008;%%
  %46 citations counted in INSPIRE as of 16 Nov 2017

\bibitem{Giombi:2016hkj}
S.~Giombi and V.~Kirilin, \emph{{Anomalous dimensions in CFT with weakly broken
  higher spin symmetry}}, \href{http://dx.doi.org/10.1007/JHEP11(2016)068}{JHEP
  {\bf 11} (2016)  068},
\href{http://arxiv.org/abs/1601.01310}{{\tt arXiv:1601.01310 [hep-th]}}.
%%CITATION = ARXIV:1601.01310;%%.
\bibitem{Hellerman:2015nra}
S.~Hellerman, D.~Orlando, S.~Reffert, and M.~Watanabe, \emph{{On the CFT
  Operator Spectrum at Large Global Charge}},
  \href{http://dx.doi.org/10.1007/JHEP12(2015)071}{JHEP {\bf 12} (2015)  071},
\href{http://arxiv.org/abs/1505.01537}{{\tt arXiv:1505.01537 [hep-th]}}.
%%CITATION = ARXIV:1505.01537;%%.

\bibitem{Alday:2016njk}
L.~F. Alday, \emph{{Large Spin Perturbation Theory for Conformal Field
  Theories}}, \href{http://dx.doi.org/10.1103/PhysRevLett.119.111601}{Phys.
  Rev. Lett. {\bf 119} (2017) no.~11, 111601},
\href{http://arxiv.org/abs/1611.01500}{{\tt arXiv:1611.01500 [hep-th]}}.
%%CITATION = ARXIV:1611.01500;%%.


\bibitem{Alday:2016jfr}
  L.~F.~Alday,
  \emph{{Solving CFTs with Weakly Broken Higher Spin Symmetry}},
  \href{http://dx.doi:10.1007/JHEP10(2017)161}{JHEP {\bf 10} (2017) 161},
  \href{http://arxiv.org/abs/1612.00696}{{\tt arXiv:1612.00696 [hep-th]}}.
  %%CITATION = doi:10.1007/JHEP10(2017)161;%%
  %25 citations counted in INSPIRE as of 14 Nov 2017ITATION = ARXIV:1612.00696;%%.

\bibitem{Penedones:2015aga}
J.~Penedones, E.~Trevisani, and M.~Yamazaki, \emph{{Recursion Relations for
  Conformal Blocks}}, \href{http://dx.doi.org/10.1007/JHEP09(2016)070}{JHEP
  {\bf 09} (2016)  070},
\href{http://arxiv.org/abs/1509.00428}{{\tt arXiv:1509.00428 [hep-th]}}.
%%CITATION = ARXIV:1509.00428;%%.
\bibitem{Weinberg:2012cd}
S.~Weinberg, \emph{{Minimal fields of canonical dimensionality are free}},
  \href{http://dx.doi.org/10.1103/PhysRevD.86.105015}{Phys. Rev. {\bf D86}
  (2012)  105015},
\href{http://arxiv.org/abs/1210.3864}{{\tt arXiv:1210.3864 [hep-th]}}.
%%CITATION = ARXIV:1210.3864;%%.
\bibitem{Costa:2011mg}
  M.~S.~Costa, J.~Penedones, D.~Poland and S.~Rychkov,
  \emph{Spinning Conformal Correlators},
   \href{http://dx.doi:10.1007/JHEP11(2011)071}{JHEP {\bf 1111} (2011) 071},
  \href{http://arxiv.org/abs/1107.3554}{{\tt arXiv:1107.3554 [hep-th]}}.
  %%CITATION = doi:10.1007/JHEP11(2011)071;%%
  %183 citations counted in INSPIRE as of 14 Nov 2017

\bibitem{Dobrev:1976vr}
V.~K. Dobrev, G.~Mack, I.~T. Todorov, V.~B. Petkova, and S.~G. Petrova,
  \emph{{On the Clebsch-Gordan Expansion for the Lorentz Group in n
  Dimensions}},
\href{http://dx.doi.org/10.1016/0034-4877(76)90057-4}{Rept. Math. Phys. {\bf 9}
  (1976)  219--246}.
%%CITATION = RMHPB,9,219;%%.
%\cite{Kehrein:1994ff}

\bibitem{Kehrein:1994ff}
  S.~K.~Kehrein and F.~Wegner,
  \emph{The Structure of the spectrum of anomalous dimensions in the N vector model in (4-epsilon)-dimensions},
  \href{http://dx.doi:10.1016/0550-3213(94)90406-5}
  {Nucl.\ Phys.\ B {\bf 424} (1994) 521},\href{http://arxiv.org/abs/hep-th/9405123}{{\tt hep-th/9405123}}
  %%CITATION = doi:10.1016/0550-3213(94)90406-5;%%
  %16 citations counted in INSPIRE as of 27 Nov 2017

\bibitem{Caron-Huot:2017vep}
S.~Caron-Huot, \emph{{Analyticity in Spin in Conformal Theories}},
  \href{http://dx.doi.org/10.1007/JHEP09(2017)078}{JHEP {\bf 09} (2017)  078},
\href{http://arxiv.org/abs/1703.00278}{{\tt arXiv:1703.00278 [hep-th]}}.
%%CITATION = ARXIV:1703.00278;%%.

\bibitem{Simmons-Duffin:2017nub}
D.~Simmons-Duffin, D.~Stanford, and E.~Witten, \emph{{A spacetime derivation of
  the Lorentzian OPE inversion formula}},
\href{http://arxiv.org/abs/1711.03816}{{\tt arXiv:1711.03816 [hep-th]}}.
%%CITATION = ARXIV:1711.03816;%%.
\end{thebibliography}

\end{document}